\documentclass[mathleft
]{an}
\usepackage{graphicx}
\usepackage{times}
\begin{document}

\Pagespan{1}{4}
\Yearpublication{2010}%
\Yearsubmission{2009}%
\Month{}%
\Volume{}%
\Issue{}%

\title{V5116 Sgr: a disc-eclipsed SSS post-outburst nova?\thanks{Based on observations obtained with XMM-Newton, an ESA science mission with instruments and contributions directly funded by ESA Member States and NASA}}

\author{G. Sala\inst{1}\fnmsep\thanks{Corresponding author:
  \email{gloria.sala@upc.edu}\newline}
, M. Hernanz\inst{2}, C. Ferri\inst{2} \and  J. Greiner\inst{3}
}
\titlerunning{V5116 Sgr: a disc-eclipsed SSS post-outburst nova?}
\authorrunning{Sala, Hernanz, Ferri, Greiner}
\institute{
Departament de F\'isica i Enginyeria Nuclear (DFEN), Universitat Polit\`ecnica de Catalunya (UPC-IEEC). C. Comte d'Urgell 187, 08036 Barcelona, Spain.
\and 
Institut de Ci\`encies de l'Espai (CSIC-IEEC). Campus UAB. Facultat de Ci\`encies, 08193 Bellaterra, Spain
\and 
Max-Planck-Institut f\"ur extraterrestrische Physik. PO Box 1312. 85741 Garching bei M\"unchen, Germany}

\received{}
\accepted{}
\publonline{later}


\abstract{%
Nova V5116 Sgr 2005 No. 2, discovered on 2005 July 4, was observed with XMM-Newton in March 2007, 20 months after the optical outburst. The X-ray spectrum showed that the nova had evolved to a pure supersoft X-ray source, indicative of residual H-burning on top of the white dwarf. The X-ray light-curve shows abrupt decreases and increases of the flux by a factor 8 with a periodicity of 2.97~h, consistent with the possible orbital period of the system. The EPIC spectra are well fit with an ONe white dwarf atmosphere model, with the same temperature both in the low and the high flux periods. This rules out an intrinsic variation of the X-ray source as the origin of the flux changes, and points to a possible partial eclipse as the origin of the variable light curve. The RGS high resolution spectra support this scenario showing a number of emission features in the low flux state, which either disappear or change into absorption feature in the high flux state. 
A new XMM-Newton observation in March 2009 shows the SSS had turned off and V5116 Sgr had evolved into a weaker and harder X-ray source.}

\maketitle

\section{Introduction}

V5116 Sgr was discovered as Nova Sgr 2005b on 2005 July 4.049~UT, 
with magnitude $\sim$8.0, rising to mag 7.2 on July 5.085 \cite{lil05}. 
It was a fast nova ($t_3=39$ days) belonging to the Fe II class in the Williams (1992) 
classification (Williams et al. 2008).
The expansion velocity derived from a sharp P-Cyg profile
detected in a spectrum taken on July 5.099 was $\sim$1300~km/s.
IR spectroscopy on July 15 showed emission lines with FWHM~$\sim$2200~km/s \cite{rus05}. 
Photometric observations obtained during 13 nights in the period August-October 2006 
show the optical light curve modulated with a period of $2.9712\pm0.0024$~h \cite{dob07},
which the authors interpret as the orbital period. They propose that the  
light-curve indicates a high inclination system with an irradiation effect on the secondary star. 
The estimated distance to V5116~Sgr from the optical light curve is $11\pm3$~kpc \cite{sal08}.
A first X-ray observation with Swift/XRT (0.3--10~keV) in August 2005 yielded 
a marginal detection with 1.2($\pm$1.0) $\times10^{-3}$~cts/s \cite{nes07a}. Two years later,  
on 2007 August 7, Swift/XRT showed the nova as a SSS, 
with 0.56 $\pm$0.1~cts/s \cite{nes07b}. A first fit with a blackbody indicated 
$T\sim4.5\times10^5K$. A 35~ks Chandra spectrum obtained 
on 2007 August 28 was fit with a WD atmospheric 
model with N$_{\rm H}=4.3\times10^{21}$~cm$^{-2}$ and $T=4.65\times10^5K$ \cite{NelOri07}.

\section{The 2007 XMM-Newton observation}

V5116 Sgr (Nova Sgr 2005 No. 2) was one of the targets included in our X-ray monitoring programme
of post-outburst Galactic novae with XMM-Newton (see contribution by M. Hernanz in these proceedings). 
It was observed with XMM-Newton \cite{Jan01} on 2007 March 5, 610 days after outburst
(obs. ID: 0405600201, Sala et al. 2007a). 
The exposure times were 12.7~ks for the European Photon Imaging Cameras (EPIC) 
MOS1 and MOS2 \cite{Tur01}, 8.9~ks for the EPIC-pn \cite{Str01}, 
12.9~ks for the Reflection Grating Spectrometer, RGS \cite{Her01}, 
and 9.2~ks for the Optical Monitor, OM \cite{Mas01}, used with the U filter in place.

The observation was affected by solar flares, which produced
moderate background in the EPIC cameras for most of the exposure time. 
Fortunately enough, our target was at least a factor 10 brighter than the background.
In addition, the source spectrum is very soft and little affected by solar flares.
We therefore do not exclude any time interval of our exposures, but we pay 
special attention to the background subtraction of both spectra and light-curves.
Data were reduced using the XMM-Newton Science Analysis System 
(SAS~7.1.0). Standard procedures described in the SAS documentation 
(de la Calle \& Loiseau 2008, Snowden et al. 2008) were followed.

\subsection{A super-soft X-ray phase with an amazing light-curve}

\begin{figure}
\includegraphics[width=82mm,height=38mm]{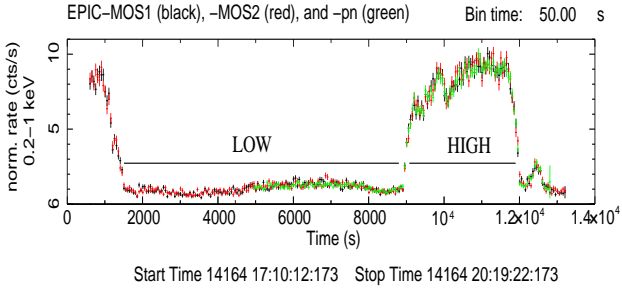}
\caption{XMM-Newton EPIC light-curve of V5116 Sgr during March 2007 observation.
Time periods for high and low states spectra extraction are indicated.}
\label{lcs}
\end{figure}

In Sala~et~al.~(2008a) we reported on the X-ray light-curve and 
broad-band spectra of the March 2007 XMM-Newton observation. 
The X-ray spectrum showed that the nova had evolved to a pure supersoft X-ray source, 
with no significant emission at energies above 1~keV.
The X-ray light-curve showed abrupt decreases and increases 
of the flux by a factor $\sim8$ (Figure \ref{lcs}), consistent with a periodicity of 2.97~h,
the orbital period suggested by Dobrotka et al. (2008).

\begin{figure}
\includegraphics[angle=0]{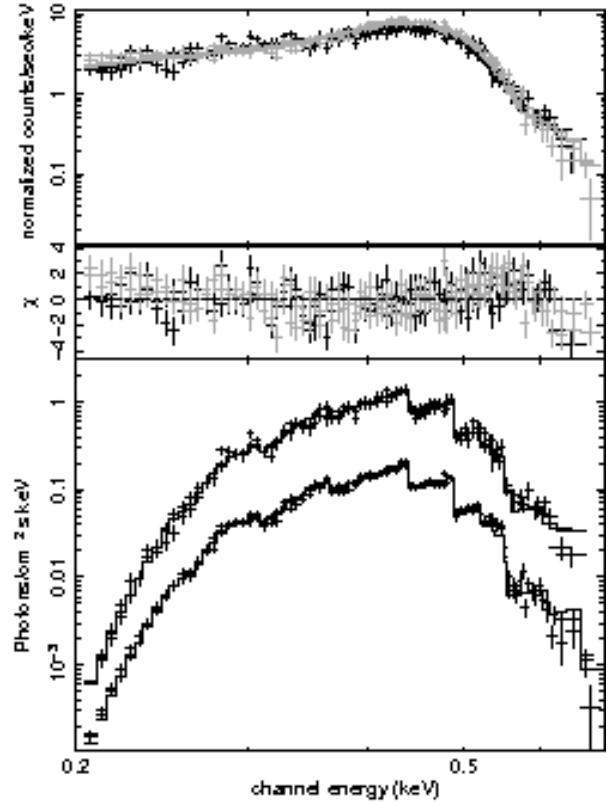}
\caption{EPIC MOS1 and MOS2 spectra of V5116 Sgr in March 2007. {\bf Upper and central panel:} 
normalised counts and residuals for the high (black) and low (grey) state spectra, 
with the best fit ONe atmosphere model (see Sala et al. 2008a for more details). 
{\bf Lower panel:} Unfolded spectra for the high and low state spectra. }
\label{spec2007}
\end{figure}

A simple blackbody model does not fit correctly the EPIC spectra, with  
$\chi^2_{\nu}>4$. In contrast, an ONe rich white dwarf atmosphere model provides 
a good fit, with 
$N_{\rm H}= 1.3(\pm0.1)\times 10^{21}\,cm^{-2}$, 
$T=6.1(\pm0.1)\times10^5\,K$, and 
$L=3.9(\pm0.8)\times10^{37}(D/10~\rm{kpc})^2{\rm erg}\,{\rm s}^{-1}$
(during the high-flux periods; Figure \ref{spec2007}). 
This is consistent with residual hydrogen burning in the white dwarf envelope. 
The white dwarf atmosphere temperature is the same both in the low and the high 
flux periods, ruling out an intrinsic variation of the X-ray source as the origin of the flux changes.
We speculate that the X-ray light-curve may result from a partial eclipse by an asymmetric 
accretion disc in a high inclination system.

The observed $N_{H}$ is consistent with the average interstellar absorption towards the source
$N_{H}=1.34\times 10^{21}cm^{-2}$  
(Kalberla et al. 2005).
Photometry at maximum of the nova outburst indicates $B-V=+0.48$ 
(Gilmore \& Kilmartin 2005), and for novae 
at maximum, intrinsic $B-V=0.23\pm0.06$ (van den Bergh \& Younger 1987). This implies 
$A_{V}=3.1\,E_{B-V}=0.8\pm0.2$. The $N_{\rm H}$ obtained from our X-ray 
spectral fits indicates $A_{V}=0.7$ 
(using $N_{\rm H}=5.9\times10^{21}E_{B-V}\,\mbox{cm}^{-2}$, Zombeck 2007), 
consistent with the value obtained from the observed colours.

\subsection{The RGS spectra}

\begin{figure}
\includegraphics[width=55mm,height=80mm,angle=90]{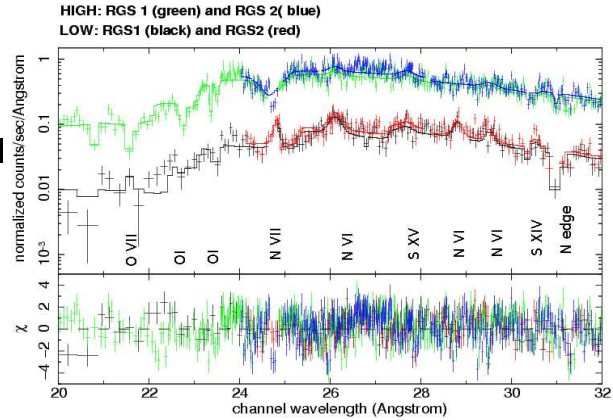}
\caption{RGS spectra of V5116 Sgr on 2007 March 13th, showing the high 
resolution spectra both in the high and low states. }
\label{spec}
\end{figure}

The light-curve of the XMM-Newton simultaneous RGS
data follows the same pattern as the EPIC data. 
We have extracted the high and low state RGS spectra running the SAS {\it rgsproc} 
RGS processing chain from the same time periods as for the EPIC spectra 
in Sala~et~al.~(2008a), as indicated in Figure~\ref{lcs}. 
The extracted RGS spectra for the high and low states are shown in
Figure~\ref{spec}.
The low state spectra show a number of clear emission lines, while many of them appear in 
absorption in the high state spectra. 

The atmosphere models used for the EPIC spectra in Sala~et~al.~(2008a) do not provide a good fit 
for the high resolution RGS spectra. In order to study the emission and absorption features
detected with the RGS without the interference of the absorption features of an atmosphere model, 
we fit the continuum between 20 and 32 $\mathrm{\AA}$ with an absorbed blackbody and determine 
the parameters of the absorption and emission lines by fitting Gaussian profiles to each feature.
The resulting best-fit model is overplotted in Figure~\ref{spec}, and 
the line parameters listed in Table~\ref{tab}.

In the 20--24~$\mathrm{\AA}$ band, the prominent absorption features in the high state correspond to 
OI~K$\alpha$ and K$\beta$, and to OVII~K$\alpha$. 
They are clearly detected in absorption in the high state spectrum.
The low statistics in this wavelength band in the low state leads to just upper limits for these lines
in the low state. Nevertheless, these upper limits for the equivalent width are compatible with the 
equivalent width determined in the high state. 

The OI~K$\alpha$ and K$\beta$ absorption lines are most probably due to the interstellar 
medium (ISM) in the V5116 Sgr line of sight. 
They are often present in the high resolution spectra of Galactic low mass X-ray binaries (Juett~et~al.~2004), 
and Yao \& Wang (2006) even used the absorption spectra of 4U 1820-303 to determine ISM oxygen and neon abundances.
The OI K$\alpha$ line at 23.52~$\mathrm{\AA}$ was also detected in the RGS spectra of the microquasar GRO J1655-40 (Sala~et~al.~2007b). In that work, the correlation between the equivalent width of the interstellar OI K$\alpha$ line and the interstellar absorbing hydrogen column was plotted using data of GRO J1655-40 and other Galactic sources, and approximated by EW(eV) = 0.5 + 2.8$\times10^{-22}$ N$_H$ (cm$^{-2}$). For V5116 Sgr, the absorbing column ($N_{\rm H}$= 1.3 ($\pm0.1) \times 10^{21}$ cm$^{-2}$) and the equivalent width of the OI K$\alpha$ line ($1.6^{+0.7}_{-2.0}$ eV) are consistent with that correlation within the uncertainties, indicating that the equivalent width of the OI K$\alpha$ line is consistent with the expected value for interstellar absorption.

The O~VII~K$\alpha$ line detected in the RGS spectra of V5116 Sgr has probably 
some ISM contribution as well.
O~VII indicates the presence of a hot absorbing gas, at T$=10^{6-7}$K,
and can be produced either by the hot ISM component, or
by some intrinsic warm absorber. O~VII has also been detected in the high resolution spectra of 
some low mass X-ray binaries, together with OI K$\alpha$ and K$\beta$.
Examples include the XMM-Newton RGS spectra of the
black-hole candidate XTE~J1817-330 (Sala~et~al.~2008b) and  the
Chandra/HETGS spectrum of GX 339-4 (Miller~et~al.~2004). 
In addition, it has even been used as a tracer of the hot component of the
Galactic ISM (Yao\& Wang 2005, 2006). Nevertheless, oxygen is an overproduced element 
in all types of classical novae explosions \cite{jh98} and there is most probably 
also an intrinsic contribution for this line originated in the nova ejecta.
 
In contrast to the Oxygen lines of the short wavelength
band, all the lines present in the  24--32~$\mathrm{\AA}$ correspond to
higher ionised species of N and S (N~VI, N~VII, S~XIV,
and S~XV) and follow a different behaviour: they show up
in emission during the low state, and are not detected or appear
in absorption (NVII) during the high state.
This pattern supports the scenario of the partial eclipse of the white dwarf: during low
flux periods, when almost 90\% of the bright X-ray source
is occulted to the observer, emission lines of photoionised
circumstellar material (which may be the nova shell itself)
become visible; in high flux periods, the bright SSS outshines the emission lines from the ejecta that were visible in the low flux period, while some of the same species
located in the circumstellar material, now in the line of sight
to the white dwarf, appear in absorption.

Highly ionised species of O, N and S have previously been detected in the ejecta of some novae
(V4743 Sgr, RS Oph; Ness et al. 2003, 2009).  Both N and O are elements overproduced 
in both CO and ONe nova explosions, while S is overproduced only in ONe novae \cite{jh98}.
So both the S detection in the RGS spectrum and the best-fit atmosphere model to the 
EPIC data being an ONe model (Sala et al. 2008a) points to V5116 Sgr hosting an ONe white dwarf.

\begin{table}
\centering
\caption{Emission and absorption lines in the RGS spectra (errors indicate 3$\sigma$ confidence if nothing else indicated.)}
\begin{tabular}{c c c c c}
\hline\hline
\noalign{\smallskip}
&  $\lambda$ (frozen)       &  		& FWHM & Eq. width \\
&  $(\rm{\AA})$    &  		& (eV)& (eV)\\
\noalign{\smallskip}
\hline
\noalign{\smallskip}
\hline
\noalign{\smallskip}

OVII	& 21.602 	&High	& $< 10$	&  $-5.3^{+2.3}_{-6.8}$	\\
		& 		&Low	& 1 (frozen)	&  $> -6$  	\\

\noalign{\smallskip}
\hline
\noalign{\smallskip}

OI~K$\beta$ 	& 22.887 	&High	& $< 13$ 	&  $-3.7^{+1.6}_{-6.5}$  \\
		&  		&Low	& 1 (frozen)	& $> -6$	\\

\noalign{\smallskip}
\hline
\noalign{\smallskip}

OI~K$\alpha$ 	& 23.507 	&High 	& $< 14$	&  $-1.6^{+0.7}_{-2.0}$  \\
		& 		&Low	& 1 (frozen)	&  $> -4$	   \\

\noalign{\smallskip}
\hline
\noalign{\smallskip}
\noalign{\smallskip}{\smallskip}
NVII		& 24.78 	&High	& $4\pm2$		&  $-6\pm2$ \\
		&		&Low	& $1.0\pm0.5(1\sigma)$	&   $5.2^{+4.5}_{-2.5}$	\\
\noalign{\smallskip}
\hline
\noalign{\smallskip}

NVI 		& 26.12 	&High	& 2 (frozen)	&  $<1$		\\
		&		&Low	& $2.0\pm0.5$	&  $4\pm2$	\\

\noalign{\smallskip}
\hline
\noalign{\smallskip}

SXV 		& 27.56 	&High	&  2 (frozen)   &  $<1.2$	\\
		&		&Low	&  2 (frozen) &  $1^{+1}_{-0.8} (1\sigma)$	\\

\noalign{\smallskip}
\hline
\noalign{\smallskip}

NVI 		&28.79		&High	& 5 (frozen)		&  $<1.3$	\\
		&		&Low	& $5\pm3 (1\sigma)$	&  $3^{+7}_{-1}$ \\

\noalign{\smallskip}
\hline
\noalign{\smallskip}

NVI 		&29.534		&High	& 2 (frozen)	&  $<1.1$	\\
		&		&Low	& $2\pm1 (1\sigma)$		&  $2^{+2}_{-1.5}$	\\

\noalign{\smallskip}
\hline
\noalign{\smallskip}

SXIV 		&30.47		&High	& 1 (frozen)   	&  $<0.4$	\\
		&		&Low	& 1 (frozen) 	&   $1.4^{+0.7}_{-1} (1\sigma)$\\

\noalign{\smallskip}
\hline
\noalign{\smallskip}
\hline
\noalign{\smallskip}
\end{tabular}
\label{tab}
\end{table}

\section{The turn-off of the SSS}

A Swift observation in June 2008 detected V5116 Sgr with a flux about
two orders of magnitude fainter than in the last Swift pointing in August 2007, indicating 
the turn-off of the Supersoft X-ray Source (Osborne 2009).

\begin{figure}
\includegraphics[width=55mm,height=80mm,angle=-90]{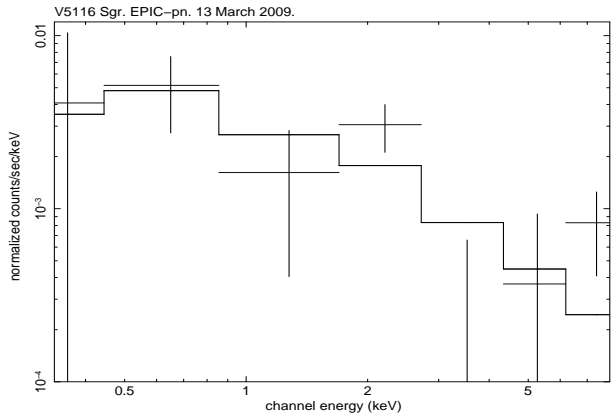}
\caption{EPIC-pn spectrum of V5116 Sgr on 2009 March 13th. }
\label{spec2009}
\end{figure}

A second XMM-Newton observation was performed on 2009 March 13, 1348 days after outburst
(Obs. ID: 0550190101, PI: Hernanz). The total EPIC exposure time was 26~ks, but after 
high-background time filtering only 14~ks of net exposure time are left. V5116~Sgr is detected
as a weak X-ray source, with a 0.2--10 keV flux in the range $(5-8)\times 10^{-14}$erg/s/cm$^2$. 
At 10~kpc, that is a luminosity of L=(3--7)$\times10^{32}(d/10$kpc)$^2$~erg/s. 
The source is too faint for a detailed spectral analysis, but the spectral distribution is certainly
harder than in the 2007 observation (see Figure~\ref{spec2009}).
The X-ray source is 2--3 times fainter than in the last Swift observation performed 9 months before.

\begin{figure}
\includegraphics[width=80mm,height=45mm,angle=0]{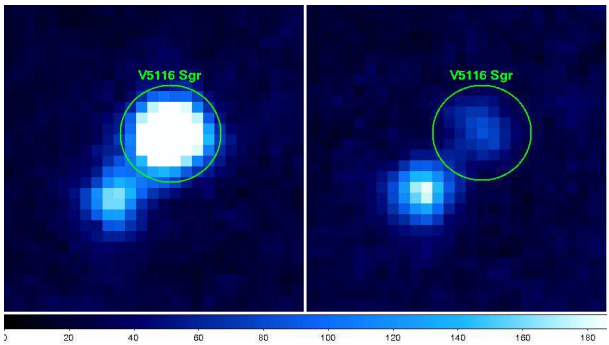}
\caption{XMM-Newton Optical Monitor U band image in March 2007 (left) and March 2009 (right).}
\label{om}
\end{figure}

The Optical Monitor was used in fast mode with U and UVW1 filters. Two sources are present in the 
fast mode window, making the analysis of the data more complex (currently underway). 
The simultaneous image mode data show the optical counterpart to V5116~Sgr fainter in the U band 
than in the 2007 observation (see Figure~\ref{om}), with its U magnitude increasing from 
$15.77\pm0.02$ in March 2007 to $18.1\pm0.1$ in March 2009. This is consistent with the 
scenario of the optical brightness in 2006 and 2007 being dominated by the irradiation of the secondary star
by the strong SSS source as proposed by Dobrotka~et~al.~(2008), and thus supports the orbital 
period determination by those authors.

\acknowledgements
This work is based on observations obtained with XMM-Newton, an ESA science mission with 
instruments and contributions directly funded by ESA Member States and NASA, 
and supported by BMWI/DLR (FKZ 50 OX 0001) and the Max-Planck Society.
This research has been funded by the MICINN grants AYA2008-04211-C02-01, AYA2007-66256,
and AYA2008-01839, and by FEDER funds. GS is supported by a Juan de la Cierva contract of the MICINN.

\end{document}